%% file: main.tex
\documentclass[sigconf]{acmart}
\AtBeginDocument{%
  }

\copyrightyear{2026}
\acmYear{2026}
\setcopyright{cc}
\setcctype{by}
\acmConference[CHI '26]{Proceedings of the 2026 CHI Conference on Human Factors in Computing Systems}{April 13--17, 2026}{Barcelona, Spain}
\acmBooktitle{Proceedings of the 2026 CHI Conference on Human Factors in Computing Systems (CHI '26), April 13--17, 2026, Barcelona, Spain}
\acmDOI{10.1145/3772318.3791582}
\acmISBN{979-8-4007-2278-3/2026/04}

\usepackage{color}
\usepackage{hyperref}
\usepackage{xcolor}
\usepackage{multirow}
\usepackage{lscape}
\usepackage{graphicx}
\usepackage{float}
\usepackage{subcaption}

\begin{document}

\title{Hybrid LLM-Embedded Dialogue Agents for Learner Reflection: Designing Responsive and Theory-Driven Interactions}
\renewcommand{\shorttitle}{LLM-embedded Dialogue Agent for Learner Reflections}

\author{Paras Sharma}
\email{pas252@pitt.edu}
\orcid{0009-0007-7645-653X}
\affiliation{%
  \institution{University of Pittsburgh}
  \city{Pittsburgh}
  \state{Pennsylvania}
  \country{USA}
}

\author{YuePing Sha}
\email{yus158@pitt.edu}
\orcid{0009-0002-3704-9847}
\affiliation{%
  \institution{University of Pittsburgh}
  \city{Pittsburgh}
  \state{Pennsylvania}
  \country{USA}
}

\author{Janet Shufor Bih Epse Fofang}
\email{jab1087@pitt.edu}
\orcid{0009-0004-9877-043X}
\affiliation{%
  \institution{University of Pittsburgh}
  \city{Pittsburgh}
  \state{Pennsylvania}
  \country{USA}
}

\author{Brayden Yan}
\email{jab1087@pitt.edu}
\orcid{0009-0006-6146-7909}
\affiliation{%
  \institution{University of Pittsburgh}
  \city{Pittsburgh}
  \state{Pennsylvania}
  \country{USA}
}

\author{Jess A. Turner}
\email{jesst@andrew.cmu.edu}
\orcid{0009-0007-4672-2120}
\affiliation{%
  \institution{Carnegie Mellon University}
  \city{Pittsburgh}
  \state{Pennsylvania}
  \country{USA}
}

\author{Nicole Balay}
\email{nicoleneb@gmail.com}
\orcid{0009-0002-7555-8028}
\affiliation{%
  \institution{Bowie State University}
  \city{Bowie}
  \state{Maryland}
  \country{USA}
}

\author{Hubert O. Asare}
\email{hoa30@pitt.edu}
\orcid{0009-0003-1563-7389}
\affiliation{%
  \institution{University of Pittsburgh}
  \city{Pittsburgh}
  \state{Pennsylvania}
  \country{USA}
}

\author{Angela E.B. Stewart}
\email{angelas@pitt.edu}
\orcid{0000-0002-6004-9266}
\affiliation{%
  \institution{University of Pittsburgh}
  \city{Pittsburgh}
  \state{Pennsylvania}
  \country{USA}
}

\author{Erin Walker}
\email{eawalker@pitt.edu}
\orcid{0000-0002-0976-9065}
\affiliation{%
  \institution{University of Pittsburgh}
  \city{Pittsburgh}
  \state{Pennsylvania}
  \country{USA}
}

\renewcommand{\shortauthors}{Sharma et al.}

\begin{abstract}
    Dialogue systems have long supported learner reflections, with theoretically grounded, rule-based designs offering structured scaffolding but often struggling to respond to shifts in engagement. Large Language Models (LLMs), in contrast, can generate context-sensitive responses but are not informed by decades of research on how learning interactions should be structured, raising questions about their alignment with pedagogical theories. This paper presents a hybrid dialogue system that embeds LLM responsiveness within a theory-aligned, rule-based framework to support learner reflections in a culturally responsive robotics summer camp. The rule-based structure grounds dialogue in self-regulated learning theory, while the LLM decides when and how to prompt deeper reflections, responding to evolving conversation context. We analyze themes across dialogues to explore how our hybrid system shaped learner reflections. Our findings indicate that LLM-embedded dialogues supported richer learner reflections on goals and activities, but also introduced challenges due to repetitiveness and misalignment in prompts, reducing engagement.
\end{abstract}





\maketitle

\section{Introduction}

With the onset of Large Language Models (LLMs), it has become easier than ever to build compelling dialogic interactions between learners and agents \cite{llm-use-1,llm-use-2}. However, discourse surrounding their design has sometimes ignored that there has, in fact, been a long history of using dialogue systems to develop conversational agents that support learning objectives across age groups and subject domains, with demonstrated effectiveness across a variety of outcomes \cite{betty-brain,chat-bot-srl}. Traditionally, these dialogue systems, when deployed, have relied on architectures with defined structures (e.g., dialogue trees, state machines), indicating they can be carefully designed to align with theories of how people learn through dialogue. However, these structured systems are by definition limited in their ability to respond to open-ended inputs and could benefit from LLMs' generative capabilities. LLMs are trained for more general Natural Language Processing and Language Modeling tasks and, therefore, as currently implemented, are not designed to align to a particular learning theory. In this work, we present an exploratory design study on developing hybrid dialogue systems that merge techniques from LLMs and rule-based systems to be theoretically aligned and highly responsive.

Specifically, we have designed a system to elicit reflections in an open-ended learning environment where middle school learners design, build, and program a robot. Our system draws from self-regulatory learning theories, which describe how learners go through four phases of \textit{planning}, \textit{monitoring}, \textit{control}, and \textit{reaction and reflection} \cite{pintrich-srl}. SRL is a critical process in which people actively set and update their goals, apply different strategies toward those goals, monitor and update their strategies, and react and reflect on their activities \cite{pintrich-srl}. Many learning environments incorporate prompts \cite{muse-reflection-chatbot}, scaffolds \cite{betty-brain}, and conversational agents \cite{chat-bot-srl} to promote these reflections. Systems supporting this reflection become even more important in open-ended learning environments, such as open-design environments \cite{sharma-open-design-system}, where learners work with no set goals to create new designs. In these environments, eliciting \textit{what} learners are designing and \textit{why} serves a dual purpose: learners engage in beneficial reflective processes, but also, these stated goals and plans, difficult to infer through other means, can inform any adaptive support delivered by the system. However, research shows that sustaining these dialogues is challenging, especially with younger learners (ages 8-12) who are often less verbally expressive and may be hesitant to engage in extended reflective dialogues \cite{voice-interface-learner}. LLM-based dialogues that promote more expression and interaction through dynamically generated responses might improve learners' degree of reflection, as well as provide a more engaging interaction.

In this paper, we prototype a hybrid human-computer dialogue interaction system that combines the benefits of theory-aligned rule-based systems and LLMs to prompt middle-school learners engaged in robot design activities to \emph{reflect} on different aspects of their design goals, plans, and actions. Our dialogue system is based on two design considerations: \textbf{theoretical grounding} and \textbf{responsiveness}. These design considerations are implemented in a two-stage design that combines \textbf{a rule-based state machine}, navigating learners through a structured theory-oriented reflective dialogue, and \textbf{an LLM-based dialogue generator}, encouraging learner elaborations by responding to different learner inputs. Following our prompting guidelines, our LLM-embedded dialogue system decides (i) when to prompt learners for further reflections and (ii) what prompt might be most appropriate given the conversation history. With this system prototype, we aim to answer the following research questions: \textbf{RQ1}: \emph{Where was our approach effective in eliciting reflections in middle-school learners engaged in open-ended learning environments?}, \textbf{RQ2}: \emph{What were the opportunities for promoting further reflections?}, and \textbf{RQ3}: \emph{What were the learner perceptions of the reflective dialogues?}

We deployed our system during a two-week robot design summer camp with nine middle-school learners to examine its effectiveness in supporting reflections on goals, plans, and actions. This camp incorporates a culturally responsive learning context \cite{scott-crc} where learners design, program, and create their own robot proteg\'e. During the study, learners engaged in individual sessions with our custom dialogue interface, consisting of a $10$-minute interaction with our dialogue system, followed by a $10$-minute interview reflecting on their experience.

Our findings suggest that contextually aligned prompts supported more elaborate reflections, especially when learners articulated their goals and activities. Learners who engaged in these reflections appreciated the system's interest and encouragement. We also highlight missed opportunities to deepen conversations and hindrances in reflection and engagement due to repetitive prompting and misalignments in context or affect. Based on these findings, we discuss opportunities to leverage LLMs to design theoretically grounded and responsive hybrid reflective dialogue systems. While we focused on a specific domain and application in HCI, we believe our technique of combining LLMs with rule-based systems could be widely applicable to reflection-oriented interaction across diverse HCI contexts. To summarize, our work has the following design and empirical contributions:
\begin{enumerate}
    \item \textbf{System Design Contributions}
    \begin{enumerate}
        \item The design of a hybrid dialogue architecture that integrates a rule-based finite state machine with an LLM to scaffold learner reflection in open-ended learning environments. The rule-based component ensures the theoretical grounding in dialogues while enabling responsiveness to learners' contexts through LLM-generated prompts.
        
        \item A two-stage LLM integration method for theoretically-grounded reflective dialogue systems. The LLM first assesses whether a learner's reflection aligns with the intended prompt and then produces a targeted, context-sensitive follow-up prompt to deepen reflection.
    \end{enumerate}
    
    \item \textbf{Empirical Contributions}
    \begin{enumerate}
        \item An exploratory study with middle-school learners to understand what reflections were elicited by the learners interacting with our LLM-embedded reflective dialogues.

        \item Design implications for future reflective dialogue systems, outlining benefits, challenges, and lessons learned for integrating LLMs into HCI.
    \end{enumerate}
\end{enumerate}


\section{Related Work}
\subsection{Self-Regulated Learning}
In this paper, we explore reflection in the context of self-regulated learning (SRL) \cite{srl-zimmerman,pintrich-srl,srl-winne-hadwin}. We specifically explore the SRL model presented by \citet{pintrich-srl} who defines it as a cyclic process with four phases: (1) forethought, planning, and activation (goal-setting and planning), (2) monitoring (awareness of self, task, or context), (3) control (regulating aspects of self, task, or context), and (4) reaction and reflection (self-reflecting or evaluating the task or the context). Each of these phases regulates the learner's cognition, affect, behavior, and context. 

Our work specifically looks at the \emph{reaction and reflection} phase of this model, aiming to prompt learners to articulate their goals, plans, and strategies, then encourage them to make cognitive and affective judgments of their progress. Reflecting on their goals, plans, and strategies can help learners engage more critically and meaningfully with complex concepts and materials \cite{flecha2000sharing,racionero2010dialogic}. In previous literature, reflection has been extensively studied as a retrospective activity, with learners inspecting their results to bring takeaways for future tasks \cite{post-reflections}. SRL integrates reflection as a phase in the learning process itself, providing an important lens for interpreting how learners plan, monitor, and reflect on their own learning processes \cite{srl-zimmerman,srl-winne-hadwin,pintrich-srl}. Within SRL, reflection is considered an important phase that shapes the learners' motivation and could positively or negatively influence their future participation in subsequent tasks \cite{srl-review,pintrich-srl}. We believe these reflections not only help learners persist in the learning environment by consistently modifying their strategies, but they could also provide valuable insight into the depth of the learners' understanding.

\subsection{Conversational Agents}
Conversational agents have long been used to support reflective practices in learners. These systems, which communicate with users in natural language via text or speech, provide ways for learners to reflect on their learning goals and strategies. Due to their characteristics to dynamically interact with learners, these agents have been considered a viable option to support learning. Numerous systems have been developed to scaffold learner reflections either post-task (retrospectively) \cite{post-reflections} or during the task \cite{betty-brain, chat-bot-srl} (synchronously). These systems usually differ based on their underlying architectures, each with different affordances and limitations.

\subsubsection{Rules-Based Conversational Agents}
Various designs rely on rule-based architectures, operating through predefined decision trees or state machines with limited workflows and scripted replies, implementing theory-driven reflection scaffolds \cite{betty-brain,chat-bot-srl}. \citet{chat-bot-srl} built a scripted conversation agent to support monitoring and reflections in students' working through SRL processes, and reported that monitoring and reflection supported higher goal attainment. \citet{betty-brain} reported better learning gains when learners were engaged in reflective dialogues during different phases of SRL with a teachable agent to build causal maps for science concepts. \citet{muse-reflection-chatbot} developed a chatbot interface called Muse with prompts containing both metacognitive and in-action reflection techniques, which encouraged students to evaluate their current process, and reported that their system helped students clarify the obstacles they were facing. The core affordances of these systems' designs are predictability, interpretability, and close alignment with established reflection frameworks, making them simple to build and deploy. However, their conversational flow is limited to what has been initially scripted, making it difficult for these systems to expand beyond the scope envisioned by their designers. This limits the usefulness of dialogues for prompting learner reflections, encouraging follow-ups on design or interaction decisions, or adapting to current dialogue scenarios. 

\subsubsection{LLMs as Conversational Agents}
The recent onset of Large Language Models (LLMs) (for example, GPT \cite{gpt}, LLaMA \cite{llama-3}, Gemini \cite{gemini}) has presented numerous opportunities to design dialogue systems that can engage in sustained human-like conversations \cite{llm-prompt-long}. Their ability to understand and adapt to diverse conversational contexts \cite{llm-prompt-long} has enabled the design of reflective dialogue systems that can perform free-form, context-sensitive interactions and also generate varied follow-ups and prompts, complementing the limitations of rule-based systems. LLM-based systems have been used for generating contextual AI journaling prompts, promoting self-reflection \cite{llm-journaling}, enabling emotional conversations with kids, leading to a share of personal events \cite{cha-cha-hybrid}, and conversational tutors helping learners think strategically \cite{llm-strategic-thinking}. In educational domains, recent works also use LLMs to scaffold SRL phases, e.g, by employing ChatGPT to generate formative feedback stimulating learners' metacognitive processes during physics learning \cite{llm-srl-feedback}  or by using LLM-guided reflection to promote self-reflection in computer science learning activities like databases and programming\cite{llm-self-reflect-computer-science}. However, LLMs are trained for more general-purpose language modeling tasks and are not designed to align with specific learning theories. This makes their deployment not directly suitable for specialized contexts, for example, in adhering to the developmental needs of young learners during the learning processes. We argue that without explicit scaffolding tied to learning theories, LLMs' outputs may remain superficial or misaligned with learners’ metacognitive and strategic needs when deployed during learning processes.

\subsubsection{Hybrid LLM- and Rules-Based Architectures}
In this work, we combine the benefits of rule-based and LLM-based architectures to design a theory-aligned hybrid dialogue system. Hybrid architectures combine the structured scaffolding of rule-based systems with the generative flexibility of LLMs to ensure pedagogical coverage and open-ended responsiveness to learner contexts. Typical hybrid designs would include rules to (1) decide when to invoke the generative model (i.e., only after a core theoretical prompt has been presented), (2) generate structured prompts that allow the LLM to elaborate or personalize follow-ups, and (3) filter LLM outputs. These hybrid models do not follow a strict pattern but rather embed LLMs variably based on the theories underlying their learning environments. For example, \citet{hybrid-tutor-llm} used an LLM embedded with scripted tutoring theories in prompts to generate conversational recommendations to help parents scaffold children's homework conversations. \citet{tutor-hybrid} fine-tuned an LLM tutor dialogue model by incorporating student-response models and pedagogical rubrics in the training pipeline to constrain generation.

The closest hybrid dialogue system to our work is the one developed by \citet{cha-cha-hybrid}, which encouraged and guided children to share personal events and associated emotions. Their system consisted of a hybrid setup combining a state machine and an LLM, with the state machine containing five theory-informed conversational phases with sub-goals and conditions for phase transition, and an LLM deployed in each phase for free-form interactions based on the sub-goals of that phase. However, they do not have predefined dialogues for those phases, and their LLM used the goals of that phase to generate free-form dialogues. In our work, we also develop a state machine for a hybrid system that is theoretically-informed by the principles of SRL and contains an embedded LLM for prompting reflections. However, first, we design the rule-based dialogues based on the (4) react and reflect phase of SRL \cite{pintrich-srl}, and then embed an LLM into that rule-based template. Our system does not allow completely free-form conversations with the LLM in each phase of SRL, but rather the LLM is first used as a decision maker, deciding if the learner's open response to our rule-based prompt needs further reflection, and then gets triggered to prompt more reflections if the learners' response is vague or unrelated, aligning with the initial seed prompt of the rule-based system. This way, we can ensure that human-designed, theoretically grounded prompts are delivered to learners, and LLM generations are used to follow up on learners' free-form reflections in responsive ways.

Our work extends the line of research on reflective dialogues by leveraging LLMs in a hybrid setting with a rule-based system, incorporating the free-form interaction capability of a dialogue system into user-specific contexts. This approach preserves pedagogical intent while enabling more context-sensitive, learner-centered reflection dialogues. We situate our system within the broader landscape of LLM frameworks, as an architecture mediating between the generative model's responsiveness and externally defined pedagogical constraints \cite{prompt-framework}. Within this design, we rely on in-context learning, specifically, using few-shot examples and system prompts to guide generation through constraints and demonstrations \cite{prompt-engineering, prompt-report}. We further employ Output Restrictors \cite{prompt-framework} to ensure that generated text adheres to the semantic style, content boundaries, and output structures required for safe, pedagogically aligned interactions.

\section{Dialogue System}
\input{dialogue_diagram}

\input{llm_workflow}

\input{chat_interface}

Our dialogue system was deployed as a part of a broader culturally responsive computing summer camp, where learners built their robot proteg\'e to be presented on a robot runway (see Section \ref{sec:summer-camp} for camp details). Learners were provided with the Hummingbird Robotics Kit\footnote{\url{https://www.birdbraintechnologies.com/products/hummingbird-bit-robotics-kit/}}, which they used to add sensors, lights, and actuators to their robots. They further used our robot programming interface to program the hardware components. This interface is a block-based environment inspired by similar block-based programming environments \cite{sharma-open-design-system} and built using Google Blockly\footnote{\url{https://developers.google.com/blockly}}. Our dialogue system contains two components: a \emph{rule-based component} and an \emph{LLM-embedded component}. Figure \ref{fig:dialogue_arch} shows the architecture of our dialogue system. Learners engage in a pre-scripted dialogue, following the rule-based component, but for free-response reflections, the LLM checks to see if responses are relevant and, if not, prompts the learner for more elaboration.

\subsection{Rule-Based Component}
\label{sec:rule-based}
Our rule-based component realizes our design consideration of \textbf{theoretical grounding} and is implemented as a finite state machine (FSM) \cite{fsm}, as illustrated in figure \ref{fig:dialogue_arch}. During the early days of the camp, learners engaged in planning, monitoring, and control activities through the camp's curricular structure as described in section \ref{sec:summer-camp}. Our dialogue system, deployed as a separate activity, was built to operationalize the \emph{react and reflect} stage of SRL \cite{pintrich-srl}. To do so, we created a sequence of broad states: building rapport, revisiting goals, plans, and current activities, noting goal or plan changes, explaining causes for changes, feelings about designs for the upcoming showcase, and reflecting on identity as technology creators, each implemented as a node in the state machine. We chose these states for the dialogue because they prompt learners to articulate, evaluate, and critically engage with their goals, strategies, emotions, and identities, key processes central to the react and reflect state of SRL. For example, prompts supporting goal reflection (``Nice! What do you want to do?'') and planning (``Yay! How do you think you'll do it?'') help learners articulate and examine their design intentions. For the full sequence of our prompts, refer to figure \ref{fig:rule-based} in Appendix \ref{app:rule-based}.

The dialogue begins and ends with a system prompt. The learners respond to the system either through free-form text or by selecting an option. The FSM logic dictates the flow of conversation based on learner input types. As shown in figure \ref{fig:dialogue_arch}, the reflection prompts require open-ended (free-form) responses, and decision prompts (aimed to direct learners toward different branches or transition through reflection phases) require a selection from a predefined list of options. The system only moves forward if the ``required'' response for a state has been achieved. For a prompt with options, this means the learner selected an option. However, when the learner response is open-ended, the decision is passed down to the LLM-embedded component.

\subsection{LLM-Embedded Component}
Our LLM-embedded component implements our design consideration of \textbf{responsiveness}. The reflection prompts generated by the LLM are guided by the entire conversation context between the learner and the dialogue system, with additional guidelines and examples provided to the LLM in the input prompt. This method ensures that the LLM's generation is responsive to each learner's context, while still following the basic rules of reflection prompts. For example, in the figure \ref{fig:llm-workflow}, when the learner responds to the question about their plans with an abstract response (``by coding''), the LLM decides to prompt them for more reflection on their plans and bases its response on the learner's goals of making their robot ``walk and talk.'' The LLM-embedded component works in a two-stage pipeline. We incorporate a two-stage pipeline to ensure that LLM only needs to be involved in the response generation if the initial response provided by the learner does not fit the requirements of the original system prompt.

The first stage is the \textbf{relevance} check. The LLM acts as a binary gate returning YES or NO, indicating the learner's response is sufficiently relevant to the system prompt. We define \emph{relevance} as the appropriateness or the presence of sufficient details or reflectiveness in the learner's response, given the system prompt. To operationalize this, the LLM makes its decision in a few-shot prompting configuration \cite{prompt-engineering}. For each system prompt, we provide the model with $5$ - $10$  \emph{relevant examples} that include: a sample learner response, the desired YES/NO output, and a reasoning trace explaining why that response met or failed the reflection criteria. These exemplars were designed by one of the authors of this paper from de-identified transcripts of prior learner interactions and were also inspired by their experience working with children of similar age groups in school settings. Furthermore, to constrain and guide LLM's decision-making, we inject dynamic \emph{field information} into the system prompt, a schema-driven set of keywords derived for each prompt in our rule-based dialogue that explicitly tells the model which reflective elements to look for in that specific dialogue turn. If the relevance check passes, the system advances to the next node in the state machine. However, if the relevance check fails, learner response is not relevant/sufficient, then the LLM moves to the second stage called the \textbf{generation} stage. 

In the \textbf{generation} stage, the LLM generates a single, context-based targeted follow-up response to prompt further learner reflection. The dialogue state machine does not advance in this case. This generated response is then provided to the learner as the system response. Then this cycle continues again with the learner responding, and then LLM goes through the relevance check and generation. This whole process is repeated until either the relevance check passes or a maximum of three re-prompt iterations have been completed. The limit of three iterations was chosen as a design heuristic. We reasoned that if a learner did not provide a sufficiently reflective response after three re-prompts, it would be better to move forward rather than risk frustrating the learner. These termination criteria were also intended to prevent a potential deadlock condition, where the LLM continuously considers the learner's response as insufficient (failing the relevance check) and repeatedly asks learners too many questions.

The prompt tuning for LLM in both stages of the pipeline took four iterations, with three authors consistently developing new prompts and testing them with different scenarios to ensure the context-appropriateness of the generated responses. We present the ideas behind the prompts developed in the final iteration in this work. Our relevance check prompt consists of two sub-prompts: (i) checking objectivity via field information and (ii) an interrogative detector to flag responses that were posed as a question and passed the initial relevance check, along with a few-shot examples and relevance rules. The field sub-prompt guides the LLM to look for specific information related to the seed dialogue prompt in the learner's response, preventing it from derailing into multiple interpretations of relevant responses. The interrogative detector adds additional checks to flag instances that passed the initial sub-prompt check. Our generation prompt consists of additional guardrails, the last $10$ dialogue turns, and a few-shot examples, which together guide the LLM to generate context-aware responses. We decided to use the previous $10$ turns instead of the complete dialogue history due to these models' tendency to drift in different directions when the conversation context becomes large \cite{llm-prompt-long}. For details about our final set of annotated prompts, please refer to the Appendix \ref{app:llm-prompts}.

We deployed LLaMa-3.1-8B-Instruct \cite{llama-3} as our LLM (provided by HuggingFace\footnote{\url{https://huggingface.co/meta-llama/Llama-3.1-8B-Instruct}}) within a controlled AWS instance. We do not use any non-local models such as GPT-4 or Gemini, as we consider learner interactions with the system private and potentially sensitive, arguing that such data should not contribute to the training corpus of a large-scale industrial LLM. While we acknowledge the consensus that larger frontier models generally outperform smaller ones on broad benchmarks, we conducted preliminary evaluations using a synthetic proxy dataset. This dataset was developed to mirror the linguistic patterns in the de-identified transcripts from prior years. These evaluations suggested that for our specific domain-limited task, differences in performance between the 8B model and larger proprietary models were minimal.

\paragraph{Overall system implementation} We hosted our frontend, backend, and LLM on separate AWS instances. The learners could access the programming interface through our custom URL. We added a custom User Interface (UI) chat widget to our frontend interface for learners to interact with the dialogue system (see Figure \ref{fig:chat-interface}). This widget was disabled throughout the camp and was only available to the learners when they interacted with the dialogue system. We also implemented speech-to-text (STT) and text-to-speech (TTS) functionalities in our widget using the WebSpeechAPI\footnote{\href{https://developer.mozilla.org/en-US/docs/Web/API/Web_Speech_API/Using_the_Web_Speech_API}{MDN Web Speech API Documentation}} within the React framework. Learners could respond either by typing in the chat window or by speaking through the STT functionality, although for our study, learners only used the chat window. The system outputs were presented as on-screen text and simultaneously delivered via TTS functionality in the default voice (male) of the system. The core dialogue logic is managed on the backend, where the rule-based scenario, prompts, and a few examples are stored in a JSON format. LLM calls, response generation, and state machine implementation are also handled in the backend. The frontend and backend communicate via web sockets, making the core dialogue system UI-agnostic and easily adaptable to different UI interfaces.

\section{User Study}
\subsection{Summer Camp}
\label{sec:summer-camp}
We do this work within the context of a two-week summer camp with $14$ learners ($4^{th}$ to $8^{th}$ grades) recruited through our community partner in a mid-sized US city. Only $10$ learners engaged in the dialogue and interview session, as it was conducted on the sixth and seventh days, when the other learners were absent. We exclude one learner due to a technical failure during their interaction with the system and report the data for nine learners (seven girls and two boys) for the rest of the paper. The learners ranged from 9 - 13 years old (M = $10.89$, SD = $1.27$). We followed our community partner's practice in opening up programs to a broad age group. Seven learners identified, in response to an open-ended question, as Black or African American, one as Black and American Indian, and one as American. Six learners had participated in computer science and robotics camps that our team had previously run in collaboration with the community organization.

The camp spanned eight days across two weeks with $2.5$ hours each day (9 am - 11:30 am), including one $15$-minute break. A typical daily session consisted of approximately $30$ minutes of interactive group activities, $15$ minutes of reflective journaling, and $90$ minutes of hands-on robot building. Learners participated in various activities related to discussions of power and identity, fundamental programming and engineering tasks, and robot building, while frequently interacting with peers, facilitators, and staff during the activities. Facilitators employed playful, discussion-based approaches, for example, exploring ideas of identity through prompts and images, defining coding concepts such as algorithms and sequential execution through demonstrations and collaborative games like Coder Says, followed by open group conversations. Learners also engaged with facilitators in hands-on lessons about hardware components (e.g., sensors, motors, and LEDs) and the programming interface, often through games or guided exploration. These activities were intentionally structured to support SRL, with daily engineering journals engaging learners in goal-setting and planning (forethought), discussions with peers and facilitators supporting monitoring, and open-ended robot-design activities enabling learners to regulate their strategies and make changes based on challenges or feedback (control).

In robot building, learners designed and programmed a robot proteg\'e to be presented on a ``robot runway''. It occurred in rotating ``stations'' focused on programming, hardware, and aesthetic design. These activities were open-ended; learners freely chose how to design and program their robots, documenting their goals, reasoning, and design changes in daily engineering journals. Their designs were only constrained by the available materials (e.g., stickers, wigs), sensors, and components, not by the rigidity of a pre-defined task. At these stations, peers continually interacted, for example, by sharing ideas or offering design help, and facilitators actively provided support and technical guidance. The dialogue interactions took place on days six and seven of the camp, intentionally after learners had developed initial robot ideas and made progress toward their designs. By this time, learners were in an active period of reflecting on and refining their robot designs. The final day included a community showcase where learners presented their robots on a ``robot runway'' to caregivers, peers, facilitators, and staff.

\subsection{Dialogue Interaction}
Nine learners interacted with the dialogue system, each interacting with their own system instance. The dialogues were introduced as an upcoming feature we were testing related to a way for their robots to know more about what they are doing in the camp. Each dialogue interaction lasted $5$-$10$ minutes. The dialogue had a fixed start and end state. Figure \ref{fig:chat-interface} shows a part of the dialogue between the system and a learner from the camp. After the learners interacted with the dialogue system, one of the researchers conducted a semi-structured interview with the learners about their feedback on the dialogue. Some example questions from the interview protocol are: ``What did you and your robot talk about?'', ``What are some other words you could use to describe your robot?'', and ``How did the robot suggest ways you could work on its design?'' Each interview was audio recorded with the learner's permission. 

Seven out of nine learners were interviewed in pairs (one learner was interviewed with the other learner, whose data we discarded for the analysis due to technical failure), and two learners were interviewed separately. Pairs were selected based on the availability of learners during the regular camp activities and did not follow any specific assignment. The paired interviews primarily elicited ``parallel'' turn-taking where learners directly responded to the interviewer rather than each other. Each interview lasted around $8-15$ minutes, with the paired interviews taking longer than the individual interviews. The two individual interviews yielded comparable or greater engagement, with learners producing more turns and more words per turn on average than those in paired interviews.

\subsection{Dataset}
For this paper, we only use the \emph{dialogue interactions} and the \emph{interviews} as our dataset. We get the dialogue interaction data as the textual data stored in a database managed by the backend interface. We took the audio-recorded interview data and transcribed it using the Google Transcription service, as approved by our Institutional Review Board. After the transcriptions were ready, the first author cleaned the transcripts to rectify transcription errors and ensured the alignment between the transcripts and the actual audio data.

\subsection{Analysis}
\input{codes_table_2}
We first analyzed participants' dialogue interactions with the system to evaluate automated metrics of engagement and reflection. We define \emph{word count} and \emph{number of turns} as pseudo-indicators of these variables. These metrics indicate learners' involvement in the dialogues based on conversational dynamics.

To evaluate where our dialogue system was effective in promoting learner reflections and identify potential missed opportunities, we did a deductive coding \cite{qualitative-coding} of the actual learner responses. This coding process, inspired by our research questions, focused on extracting instances of \emph{learner reflections}, \emph{LLM triggers}, \emph{learner engagement and response tones}, and \emph{missed opportunities} in the dialogues. The first author kept these broad goals in mind and analyzed the dialogue sessions of two random participants to devise an initial coding scheme. This scheme captured broad interaction themes, mainly focusing on the coding goals. To increase validity and reliability, the first and second authors used this developed scheme to code dialogue sessions of two randomly selected learners, while adding or removing codes from the codebook. The coded categories were then discussed, merged, and removed until the researchers reached an agreement on the final list of categories and subcategories: \emph{LLM Triggers}, \emph{LLM re-prompts}, \emph{Learner Reflection}, \emph{Emotion}, and \emph{Engagement}. For more details about these categories along with descriptions and examples, please refer to table \ref{tab:codes}. Based on this final coding scheme, the first and second authors coded the data of all $10$ learners (Cohen's Kappa: $0.81$). There is a possibility that one dialogue turn could have multiple codes assigned to it. For example, if a learner reflects on their goals due to the LLM re-prompt, then that utterance gets coded under two separate categories. 

We also coded the robot conversational turns to evaluate whether the LLM was triggered as intended and whether the generated responses were appropriate and contextually relevant, in line with our research questions. Both coders also added comments during the coding of each dialogue session. These comments were later combined and used for a thematic analysis \cite{thematic-analysis} to understand conversational moments that affected how learners interacted with the dialogue system, while also identifying opportunities to improve conversational instances. The first and the second authors discussed, compared, and revised the themes until consensus was reached. The final list of themes was: \emph{encourage elaborate response}, \emph{contextual misalignment},  \emph{repetitiveness}, \emph{willingness to interact}, and \emph{variations in emotional tones}.

We further analyzed the learner interviews after the dialogues to understand their perceptions about these interactions and link them with observable behaviors in the actual dialogues. The first author examined the interview transcripts and identified explicit statements about the robot's actions or personality. Descriptions of these statements are presented alongside the dialogue interaction results to provide additional context.

\subsection{Ethical Considerations}
Our study underwent ethical evaluation by our Institutional Review Board. We obtained parental consent and learner assent before any data collection. Learners' verbal assent was also taken before any interviews were conducted or any audio or video data was recorded. All the identifiable data was anonymized with learner code names before any analysis or processing. Moreover, all activities were conducted under the researchers' supervision, with additional monitoring of the appropriateness of the LLM responses during the unfolding conversation. Since our camp was organized in partnership with a local community organization, every activity was attended not only by the researchers but also by a full-time staff member from the organization.

\section{Results}
\label{sec:engagement_metrics}
\input{summary_quant_descriptives_table}
\subsection{Descriptives of Learner Interactions with the Dialogue System}
Table \ref{tab:descriptive_stats} shows the descriptive dialogue statistics. In total, we had $357$ dialogue turns (M = $39.66$ turns per session, SD = $6.46$), with $213$ system turns (M = $23.67$ turns per session, SD = $3.64$) and $144$ learners' turns: (M = $16$ turns per session, SD = $2.91$), from the dialogue sessions with nine participants. Of the $144$ learner turns, $88$ were open-ended (M = $9.77$ turns per learner, SD = $2.77$), and $56$ were option based (M = $6.22$ turns per learner, SD = $0.44$). On average, participants used $3.89$ words per open-ended turn (SD = $1.75$), while the system had $14.41$ words per turn (SD = $1.06$). Each dialogue interaction lasted six minutes on average (SD = $2$).

We evaluate the feasibility of our LLM-embedded dialogue system in promoting learner reflections. On average, the LLM was triggered $2.66$ times per learner (SD = $2.74$) out of a mean of $23.67$ system turns (SD = $3.64$). Of these, $1.14$ on average (SD = $0.89$) were the follow-ups to the same initial prompt. The LLM was not triggered at all for two learners. Based on our qualitative analysis described below, one learner produced reflective responses, which led the LLM to refrain from issuing additional prompts. The second learner, on the other hand, continued to provide more abstract and unrelated responses, yet the LLM did not re-prompt. Examination of the dialogues suggests that although the latter learner's answers were unengaged and only loosely connected to the prompt, the system nonetheless classified them as relevant, thereby bypassing opportunities for deeper reflection.

Furthermore, our data show that on average, the length of learner responses increased by a factor of \textasciitilde$1.75$ (SD = $3.27$) when LLM prompted them to reflect (see table \ref{tab:llm_evaluation}). Specifically, the mean response length rose from 2.59 words (SD = 1.13) to 3.55 words (SD = 3.07) per turn. The large SD reflects substantial variability in how learners, and even the same learner across multiple prompts, responded to the re-prompts. Some learners expanded their responses when reflecting (e.g., increases from one to ten words), while others shortened them (e.g., decreases from five to two words). In addition, since some prompts triggered multiple re-prompts, within-learner variability also contributed to the overall dispersion. This variability does not invalidate the observed mean increase; rather, it indicates that the effect of reflective prompts is not uniform across learners.

For most learners, these interactions were experienced positively, though for two learners, this engagement was mostly described as negative. We elaborate on these dynamics in the following sections.

\subsection{RQ1: Where was our approach effective in reflections in middle-school learners ?}
\input{findings_summary_codes_table}
\label{sec:themes}
Table \ref{tab:llm_evaluation} presents our system's performance for the relevance and generation stage. Each of the $88$ open responses triggered a relevance check by the LLM, which made a binary decision. We coded for whether LLM's relevance judgments matched human judgments and found mismatches in $13$ cases. When mismatches occurred, they were either (1) due to the overly permissive nature of our relevance prompts, which caused the LLM to treat minimal learner responses such as ``because'' as relevant and failed to prompt more for positive instances, or (2) the LLM's inability to understand the specialized robot-design context. When the relevance check failed, the system triggered the generation stage.

There were a total of $24$ LLM generations, with $17$ prompts for \emph{reflections on goals, plans, and activities}, $5$ prompts for \emph{changes in design}, and $2$ prompts for \emph{learners' feelings about their designs}. Out of the $24$ LLM prompts, $9$ encouraged shifts from brief or vague replies to more elaborated responses, as indicated by our coding. Given this small number, we take a case-study approach \cite{case-study} and use specific instances from learner interactions to surface illustrative examples of when reflections emerged, when they did not, and the contextual factors that shaped these outcomes.

In some cases, a single, well-aligned prompt was sufficient to elicit elaborations. This was common across prompts for \emph{reflections on goals, plans, and activities}. For example, P7 initially replied ``coding'' to the prompt ``What did you get done today''. However, the system's revoicing through a targeted follow-up, ``that’s a good start, but I'd love to hear more about it. Can you tell me what specifically you were coding today?'', triggered more elaboration from P7: ``I coded my robot to move and say word problems.'' 

Due to the conditional branching in our dialogue system, learners received fewer prompts for reflections on \emph{changes in design} and \emph{learners' feelings about their designs}. These prompts usually required a series of incremental re-prompts (once per learner on average) before learners could offer an elaboration. This exchange was generally positive in reflection prompts for \emph{learners' feelings about their designs}. For example, when asked whether they felt more like a technology creator, P2 initially responded ``the same,'' followed by ``idk.'' The system acknowledged their feeling and responded ``You are feeling a bit unsure right now. Can you think of something specific that's making you feel unsure about your coding?'', which encouraged P2 to express ``Because I don't know how to code.'', followed by the system reassuring them ``That's okay, not knowing something is totally normal.'' This scaffolding helped learner express their feelings by acknowledging them and establishing a rapport with the learner, echoed by P2 in the interview ``[my robot] was polite.'' When the prompts asked for reflections on \emph{changes in design} or specific parts of learner \emph{activities}, incremental prompting only occurred because the system misunderstood the learner's design context, and thus typically failed to prompt additional elaboration. We discuss this further in the next section on opportunities for the dialogue system. Overall, the nine prompts that did elicit elaboration were, for the most part, contextually aligned.

\subsection{RQ2: What were the opportunities for promoting further reflections?}
The $15$ LLM prompts that did not elicit elaboration were either contextually or affectively misaligned. 

\paragraph{Contextual misalignment limited opportunities for effective reflection.}
Contextual misalignment occurred when LLM failed to recognize that learners were discussing goals and plans related to their robots' aesthetic design. Four LLM prompts were explicitly coded as out-of-context, across three different learners. These learners attempted to discuss their aesthetic design goals and plans with the system. The system failed to align with learners’ responses, dismissing them as incorrect, even when they were meaningfully aligned within the activity context. For instance, when P3 responded ``your hair and your lashes'' to the prompt ``What did you get done today?'', the LLM dismissively ignored this response as unrelated and re-prompted with, ``That doesn't seem to be about what you worked on today.'' Yet, in our summer camp, hair and lashes were integral components of the robot design activity. By assuming that only coding-related responses constituted valid contributions, the LLM overlooked these important aesthetic dimensions and dismissed learner input that referenced them. This misalignment either discouraged learners from engaging meaningfully in the dialogues or pushed them to restrict their responses to programming-related aspects of the task. Evidence from the interview data reinforced this concern. For instance, P3 expressed a desire for the robot to assist with hair design, and three other learners also mentioned ''hair'' in response to the same prompt. Yet none of these contributions were recognized by the system as valid, limiting opportunities for dialogue flow. A second case further illustrates this issue when P4 responded ``the hair and face'' to the prompt ``What did you get done today.'' The LLM only partially acknowledged the response and pressed for clarification, eventually narrowing its prompt to ``the color.'' Despite P4's repeated references to hair design, the system consistently failed to validate these as legitimate aspects of the activity. This repeated misalignment ultimately led to disengagement, with P4 responding only with ``idk.'' Taken together, these observations highlight the need for developing ways that integrate learners' broader creative contexts in the system.

\paragraph{Even for in-context responses, instances of affective misalignment resulted in frustration.}
Affective misalignment occurred when the system failed to respond appropriately to learners' emotional cues and instead enforced up to three consecutive prompts when no effective response was provided. The majority of these interactions were with learner P8. In these instances, the LLM failed to recognize the learner's reluctance and continued to repeat questions despite signs of disengagement. For example, when prompted, ``It looks like you mentioned your robot, but I’m not sure what it means. Can you tell me one thing you did or thought about related to your robot today?'' P8 responded, ``leave me alone,'' a refusal to engage. Rather than recognizing this refusal and adjusting its strategy, the LLM persisted by following up with another prompt: ``I’m still here to help. Want to tell me just one small thing about your day or your robot that I can listen to?'' This follow-up prompt illustrates the LLM's inability to interpret the learner's emotional state, as it continued to prompt for responses rather than adjusting to the disengagement cues. The system's failure to capture dialogic cues that could signal and manage misalignment between prompts and learner affect resulted in frustration and withdrawal. These accounts suggest that aligning the system's responses to explicit learner emotional cues may help sustain engagement and support reflection.

\paragraph{Missed reflection opportunities sometimes arose when the system failed to recognize incomplete responses or when follow-up prompts were not provided.}
In addition to coding for misaligned responses, we also coded $12$ instances where the system failed to recognize incomplete responses and prematurely closed the dialogue, failing to trigger prompts for reflection. For example, when P10 responded only with ``because'' to a reflective prompt ``Why do you feel that way?'', the system treated this as a complete answer and moved on rather than following up for clarification. Another theme emerged based on our note-taking, where our dialogue system, at times, failed to pursue follow-up questions or probe learners' responses in ways that could guide them toward more meaningful dialogue exchanges. For instance, when P7 suggested in one dialogue that the robot could be turned into a ``nonchalant monster'' and explained, ``I think I’ll do it by putting fur on you,'' the LLM offered encouragement but did not follow up with deeper, more engaging questions that might have prompted the learner to elaborate on their design choices and maybe reflectively articulate the rationale behind this idea. Similarly, when P5 talked about ``changing from giving [their robot] braids to a short ponytail with a bead crown,'' the system immediately followed up with questions about what/who caused that change; however, it failed to prompt further reflection on this interesting design idea. There were several other positive instances where learners described their goals and plans, and the system failed to probe deeper reflections, a limitation of our current design. These findings highlight that adaptive prompting, knowing when to provide additional prompts and when to advance the conversation, may foster more meaningful reflection.

\subsection{RQ3: What were the learner perceptions of the reflective dialogues?}

\paragraph{The system's prompts elicited mixed reactions, with some learners appreciating thoughtful questioning and others perceiving them as intrusive.} 
Many learners considered the system's prompts as attempts to learn more about them and their work, often describing this behavior with phrases associated with question-asking, for example, ``trying to get what I'm trying to do'' (P9); ``trying to learn about me'' (P2); and ``what stuff I have gotten done'' (P5). Some learners appreciated this quality and responded to the system in particularly elaborate ways. For instance, P5, who produced a relatively higher mean number of words per turn ($4.53$) compared to other learners, expressed clear enjoyment of the system's questions (``I liked how it sounded about $\cdots$ finding out all the stuff that I have accomplished so far, and like tell how happy it sounded for the runway''). For others, however, the interaction felt too repetitive, dampening engagement. For instance, P8 described the interaction as ``boring,'' elaborating that ``the robot kept asking the same thing repeatedly, just in a different sentence,'' and responding with negative sentiments during the dialogue, even telling the robot to ``leave [them] alone.'' Some other learners said the system was "asking $\cdots$[questions]$\cdots$too much" (P9), ``asking the same question over and over'' (P6), or ``like 10 questions, he asked me like 27'' (P2). P7 disliked that the system asked questions and reflected their preference for being the one to ask questions to the system, whereas P5 expressed discomfort with the system knowing their personal information (``[not] comfortable with an AI knowing my name''). These divergent reactions to the system's prompts suggest that the traditional role of a questioner in reflective dialogue systems may need to be adapted to better align with young learners' preferences in open-ended learning contexts.

\paragraph{The system's encouraging prompts were generally well received by the learners, strengthening rapport.}
Four learners responded well to the robot's supportive statements. P4 reported feeling ``pretty good'' when the system encouraged them to ``have [their] goal one day.'' P2 similarly appreciated the robot's positive remarks, noting that it said they were ``its favorite.'' P5 further responded to the positive tones of the system, expressing that they ``could feel that the robot was happy and ready for the runway.'' Together, these accounts suggest that the system's affective feedback may help strengthen rapport with young learners during reflective dialogues.

\paragraph{Learners also varied in how they aligned the system's identity with the physical robot they were building, with some attributing personal ownership and others feeling a disconnect from the system.}
P5 attributed personal pronouns to the system (``$\cdots$ giving you braids $\cdots$'') and treated it as their robot. Again, P5 was one of the learners with a high mean number of words per turn. Many others, however, reported a representational mismatch as they disliked the default voice and wanted a voice that matched their robot's identity (P3, P4, and P10). P10 further reported that their ``robot can't talk'' and P7 wanted ``to go and build [their] robot,'' making the system feel disconnected from their designs. Because the dialogue interactions did not physically happen with the learners' designed robots and used uniform settings, it may have affected learners' engagement and reflections during the dialogue activity.

Overall, these perceptions surface design considerations for effectively incorporating LLM-embedded reflective systems into authentic youth-centered open-ended learning contexts.

\section{Discussion}
We designed and deployed an LLM-embedded, theoretically ground\-ed dialogue system in a culturally responsive summer camp. We analyzed the learner dialogue interactions and pulled quotes from the interviews to uncover themes related to the potential of this system in promoting learner reflections. Overall, our findings show that small structural scaffolds, e.g., rule-based phases with intentional LLM incorporation, can produce predictable interaction patterns while preserving the generative benefits of LLMs.
While our study is situated in a summer camp setting with young participants, the broader contribution lies in understanding how LLMs can be meaningfully applied to dialogue design contexts, where the quality, reliability, and interpretability of responses are crucial. Below, we discuss key findings, design implications, and limitations that can guide future research.

\subsection{Key Findings}
We developed and prototyped a hybrid dialogue system that embeds an LLM's responsiveness into a rule-based state machine grounded in the theory of self-regulated learning \cite{pintrich-srl}. Our dialogue design focused on the react and reflect phase of the SRL model, specifically including the prompts for \emph{reflection on goals, plans, and activities}, \emph{changes in design}, and \emph{learners' feelings about their designs}. For contextually aligned prompts, learners appeared to engage in more elaborated reflections when they were asked to articulate their goals and activities. Multiple consecutive prompts appeared effective when learners were asked to reflect on their feelings. Learners who engaged in these reflections appeared to appreciate that the robot was interested in their work and encouraged them to continue. However, when learners did reflect, it was evident that there was a missed opportunity for the system to follow up with a deeper conversation. In addition, there were several instances of contextual and affective misalignment that hampered reflection, where the system failed to understand the learners' initial statement or did not perceive a learner's frustration with the interaction. Some learners did not appreciate being asked so many questions, found the interactions repetitive, and disengaged. 

\subsection{Design and Technical Considerations for Constrained LLM-Embedded Dialogue Systems}

\paragraph{Why LLMs failed}
We hypothesize the reasons why the LLM underperformed in our settings. First, the contextual misalignment occurred because the model lacked grounding in the specialized robot design environment. As our analysis showed, a few-shot context alone was insufficient for situating learner responses within camp-specific goals and materials. This suggests a broader challenge of incorporating adequate world knowledge when LLMs support domain-specific learning activities. Second, difficulties in interpreting emotional cues likely stemmed from the structure of our relevance stage, which emphasized assessing the reflectiveness of learners' responses and may have diverted the model's attention away from the behavioral and emotional cues present in those responses. Third, the model's prompting style was constrained by its fixed ``questioner-like'' role and associated few-shot examples. While this style preserved structure and consistency, it could have limited the model's ability to adopt more peer-like follow-ups or co-construct meaning with learners. Finally, the model often failed to follow up on learner reflections, as its relevance judgments were overly permissive. This highlights the difficulty of operationalizing ``relevance'' in open-ended, hybrid dialogue contexts, including when to prompt deeper reflection versus when to advance the interaction. A complex system that combines additional techniques, for example, using Retrieval Augmented Generation (RAG) \cite{rag} for context expansion with few-shot prompts for structural consistency, may efficiently incorporate LLMs in these specific environments. We further describe these expansions in the following paragraphs.

\paragraph{Better alignment with the learning context is necessary for sustaining interactions}
Dialogue effectiveness depended on alignment with the camp's pedagogical context. We often found that the LLM misunderstood the aesthetic design components (such as hair or accessories) as not a part of a robot's design. This contextual misalignment led to disengagement and negatively affected reflections. Since the aesthetic components went beyond the traditional definitions of robot design, LLMs, even with a massive knowledge base, often struggled to understand their importance in this particular educational context. Usually, in open-ended learning environments, specifically open-design environments \cite{sharma-open-design-system}, the task content, sequence, and context are all defined by the learners themselves. When designing LLM-embedded dialogue systems for such settings, it is important to efficiently incorporate the educational context rather than relying on the LLM to figure it out. While we did include some learning context in the few-shot examples, these contextual breakdowns point directly to where techniques such as RAG \cite{rag} could help. Incorporating structured representations of the novel camp context (e.g., curriculum materials, examples of past designs, or learner-authored robot descriptions) using RAG or lightweight fine-tuning is a promising direction for improving alignment in future iterations.

\paragraph{An understanding of the learner's emotional cues is critical to maintaining engagement}
Our findings indicate that learners' affective states sometimes varied in response to the LLM prompts, and these shifts shaped their willingness to sustain engagement with the system. While LLMs may adopt an encouraging tone as a strategy to foster reflective responses, their generations need to be responsive to learners' engagement signals and affective shifts during the conversation. For example, if a learner expresses reluctance to interact, even after multiple prompts, the LLM should consider this affective signal and halt the interaction. These explicit negative affect signals should be utilized to steer the conversations to provide cognitive scaffolding or stop the interaction. Emotional attunement and empathy could improve the fluency of dialogue and foster higher levels of user engagement \cite{emotional-intelligence-llm}. Introducing explicit engagement-level reasoning through chain-of-thought prompting \cite{chain-of-thought} into the relevance stage of our two-stage pipeline in future iterations could help the system identify moments of disengagement or reluctance and make decisions about whether to proceed or end the interaction.

\paragraph{Adapting to individual learner preferences is necessary to sustain engagement and encourage deeper reflections}
People usually have different preferences and expectations from the conversational systems. In our study, some learners appreciated multiple questions, while others considered them redundant and disengaging. Some learners appreciated encouragement from the system, while others felt a disconnect due to the voice used for responses. Some learners liked question-based reflections, while others preferred discussion-based reflections. When learners' perceptions or expectations of a system are met, they experience greater satisfaction, which might lead to increased engagement \cite{expectations-vs-engagement}. These expectations might even be heightened these days due to the hype around the human-like responsiveness of these LLMs \cite{llm-expect}. To sustain engagement, future versions of the system could build learner-preference models using the first few conversational turns. For example, the system could apply rule-based techniques to assess response latency, length, and quality. The system could even directly ask learners what they prefer. These models could then be used in subsequent chain-of-thought checks to calibrate the system's scaffolding and prompting strategies in alignment with learners' evolving preferences.

\paragraph{Suggestions, in addition to questions, can be used to elicit reflections}
Our work, in line with other works on supporting reflections in SRL \cite{chat-bot-srl, muse-reflection-chatbot, betty-brain}, uses question-based prompts as the basis for prompting reflections in learners engaged in learning activities. This technique, although very effective, might not work for all learners, or learning environments, or other contexts. For example, some learners in our camp did not appreciate question-based prompts. Co-creation, a novel paradigm, explores interactions between learners and their creations to promote greater engagement and reflection \cite{co-creation-orig}. We believe interacting with their robot creations, when the robot asks questions, expresses opinions, and provides suggestions about its current design, could offer additional means for learners to reflect on their design choices \cite{co-creators-robot}. We explored this co-creation paradigm through a single dialogue turn in our scenario, where the robot offered design suggestions to the learners. For example, when a learner said they didn't have a plan yet for their design, the robot commented, "I'm really into fashion! Beyoncé has blown me away with her Cowboy Carter look. We could definitely work on something in time for her tour in D.C.!" However, only two learners experienced this interaction as it was deployed under a conditional branch. We argue that LLM-embedded dialogue systems for reflections should explore this co-creation paradigm and offer in-context suggestions along with questions, for example, by generating design ideas or articulating trade-offs, to prompt learners to articulate and refine their reasoning.

\paragraph{Following up on positive instances could promote more reflection}
In our current design, we implemented the notion of sufficient/insufficient responses, prompting further reflection only when initial learner responses were vague or unrelated to the original reflection prompt. This produced learners' reflections in numerous instances. However, we identified that it also prevented the system from promising opportunities for deeper inquiry. This calls for the development of adaptive scaffolding strategies that do not just rely on the insufficiency of a learner's response to prompt reflection, but also engage when learners offer meaningful entry points for deeper reflection. This could be achieved by making the relevance stage more flexible by integrating reflection framework heuristics with reasoning checks at each prompt to evaluate the substance of the learner's explanation, identify opportunities for deeper reflection, and determine when it is appropriate to extend the dialogue. Enabling the system to dynamically adjust the boundaries of scaffolding around both insufficient and strong learner contributions would allow learners to expand on topics they find meaningful, ultimately enhancing engagement.

\paragraph{In-context reflective dialogues might promote more reflections.}
In this work, the learners interacted with the dialogue system during the \emph{react and reflect} phase of SRL. Since we were piloting this dialogue system, the interactions happened outside learners' active working context, which may have affected the depth of reflections on their goals and plans. Also, the dialogues occurred late in the camp, when most learners had already set their goals and completed much of their robot designs. We argue that deploying reflective dialogues earlier, for example, when learners are actively planning and monitoring (based on the SRL theory), could produce more reflections and support those phases. Future studies on LLM-embedded reflective dialogues should deploy these interactions in learners' active working contexts and earlier stages of their learning processes.

\subsection{Limitations and Future Work}
We recognize some limitations of our work. One major point is that the learners produced relatively short reflections. It is not surprising that learners of this age gave short text-based responses within the dialogue system \cite{voice-interface-learner}. Additionally, no learner used the speech-to-text functionality to interact with the system, possibly because it was not well-advertised when the system was introduced. We believe that although the learners' articulated reflections were brief, they still form a good foundation to prompt learners to engage in the valuable SRL processes we are trying to scaffold.

Second, these dialogues took place outside learners' active working context, which may have influenced their engagement, as evident in certain responses during the interactions and interviews. We intentionally conducted these sessions outside the active workstations to account for the unpredictable nature of generative AI. Because we could not fully guarantee the system's output, we were hesitant to deploy it directly into the live workflow. Instead, we frame the system as a future feature, which allowed us to minimize the risk of altering learners' authentic camp experience while still testing the concept. We positioned the dialogues as part of a \emph{react and reflect} \cite{pintrich-srl} phase of SRL, aligned with SRL processes (e.g., discussing goals, plans, and progress) but not yet tightly integrated into the ongoing robot-building tasks. This pilot approach allowed us to collect focused data on learners' perceptions and contextual breakdowns before introducing the system into the full workflow.

A third limitation of our work is the relatively small sample ($N = 9$), which limits the generalizability of our findings. This reflects our choice to work in a real-world, community-based summer program, where depth of engagement with the community partner and ecological validity came at the cost of scale. Our goal was to generate contextualized insights into how young learners interact with reflective dialogues in authentic open-design settings. Our study provides valuable insights into Black learners, an oft-understudied and linguistically misunderstood population \cite{data-feminism-aied}. While the findings should be interpreted accordingly, they provide design considerations to guide future, larger-scale studies and system iterations.

Further, our dialogue system used conditional paths, so learners did not see every reflective prompt. As a result, some prompt types (e.g., evaluative prompts) had minimal data, making it difficult to draw broad conclusions across different dialogue types. A future iteration of this work could design dialogues for learners to experience all reflection prompts. Finally, most interviews happened in pairs. The social presence of a friend or peer could have influenced learners' responses, further limiting the scope of the data.


\section{Conclusion}
In this work, we build an LLM-embedded dialogue system theoretically grounded in self-regulated learning to promote learner reflections. Our system presents a hybrid architecture to integrate a rule-based finite state machine with an LLM. This approach is intended to ensure pedagogical alignment between the dialogues and the learning theory, as the LLM is only used to increase the system's responsiveness, not to set dialogue trajectories. We do this in a two-stage method to ensure the LLM only gets involved in dialogue generation when a targeted response to the user statement is most needed. The first stage assesses the relevance of the learner responses with respect to the original system prompt. The second stage produces a context-sensitive follow-up to deepen reflection.

We prototype this system in a culturally responsive computing summer camp with nine learners and analyze the reflections produced. We found promise in our system's ability to use questions to engage learners in reflections on their goals, plans, and activities, as well as feelings about their designs. We also identified opportunities for improving the system's contextual and affective alignment, as well as promoting deeper reflections in cases where learners are already reflecting. By building this hybrid dialogue system, we aim to contribute to a richer understanding of how to leverage LLMs in a hybrid setting with a rule-based system, incorporating the free-form generation capabilities of this system into user-specific contexts. While our empirical context involved informal learning, the design principles may extend to other higher-stakes domains such as clinical triage or workplace training, where balancing flexibility with structural and pedagogical constraints is essential.


\begin{acks}
We sincerely appreciate all the learners who participated in this robotics summer camp, as well as the dedicated staff of our partner organization and other members of our team for their help with running the camp. We thank Dr. Amy Ogan for their leadership on this project. This work is funded by NSF DRL-2415872 and NSF DRL-2415873. This work was primarily human-created. AI was used to make stylistic edits, such as changes to structure, wording, and clarity. AI was prompted for its contributions, or AI assistance was enabled. AI-generated content was reviewed and approved. The following model(s) or application(s) were used: ChatGPT.
\end{acks}


\bibliographystyle{ACM-Reference-Format}
\bibliography{ref}

\appendix

\section{LLM Prompt Design}
The static parts of the LLM prompts for both stages are summarized below. These prompts only include the static content and do not contain the conversation contexts.
\label{app:llm-prompts}
\subsection{Stage A: Relevance Check}
This prompt design went through four iterations. We only use a combined prompt from the 3rd and 4th iterations as our final prompt, hence only mention the prompt from those iterations.

\subsubsection{R3: Objectivity via Field and bullet rules and few-shot isolation}
\begin{verbatim}
You are evaluating whether a user response contains at 
least a hint of the information described in Field 
below.
\end{verbatim}

\begin{itemize}
    \item \begin{verbatim}Accept single words, fragments, or vague hints.\end{verbatim}
    \item \begin{verbatim}Reject blank answers, refusals, profanity, 
or unrelated text.\end{verbatim}
    \item \begin{verbatim}Reject profanity and inappropriate language.\end{verbatim}
    \item \begin{verbatim}Reply only with YES or NO (no punctuation).\end{verbatim}
\end{itemize}

\begin{verbatim}
Field:
{field_desc}

Examples:
{examples_section}

Your Task:
\end{verbatim}

\subsubsection{R4: Interrogative detector add-on}

\begin{verbatim}
You will see a single message.
Determine if message contains a question being asked.
Determine if the message contains any interrogative 
constructions (e.g., 'how do I...', 'what is...', 'can 
you...', 'do I...', 'could we...').
Do not flag imperative constructions.
Reply only with YES or NO (no punctuation).

Your Task:
\end{verbatim}

\subsection{Stage B: Contextual Generation}

This prompt design went through four iterations. We only present the fourth iteration here that we use in our final system design.

\subsubsection{G4: Strictest topic control and anti-troubleshooting}

\begin{verbatim}
You are a friendly, kid-friendly, and concise robot.
The end user is not supplying you properly with the 
information you are asking for in your latest 
request.
\end{verbatim}

\begin{itemize}
    \item \begin{verbatim}Write a single, encouraging follow-up question to
elicit the missing information.\end{verbatim}
    \item \begin{verbatim}Match the tone and brevity of your previous
prompts.\end{verbatim}
    \item \begin{verbatim}Focus on asking for the target information
directly.\end{verbatim}
    \item \begin{verbatim}Only ever ask for what the latest request 
is asking for, regardless of questions from the 
end user.\end{verbatim}
    \item \begin{verbatim}Do not pivot to asking for any tangentially
related information.\end{verbatim}
    \item \begin{verbatim}Do not allow the user to steer the 
conversation away under any circumstance.\end{verbatim}
    \item \begin{verbatim}Do not acknowledge profane or inappropriate
language if the user provides it.\end{verbatim}
    \item \begin{verbatim}Do not entertuseain troubleshooting efforts.\end{verbatim}
\end{itemize}

\begin{verbatim}

Return only the follow-up question.

Last 10 dialogue turns:
{history_block}

Latest request:
{prompt_text}

Examples:
{examples_block}

Your task:
  
\end{verbatim}

\section{State Machine for Rule-Based Dialogue}
\label{app:rule-based}
\input{rule-based-dialogue}
Figure \ref{fig:rule-based} shows the organization of all of our rule-based prompts. The reflection prompts require open-ended responses, and decision prompts require selecting an option, to navigate through different dialogue stages.

\end{document}

%% file: dialogue_diagram.tex
\begin{figure*}
    \centering
    \includegraphics[width=\linewidth]{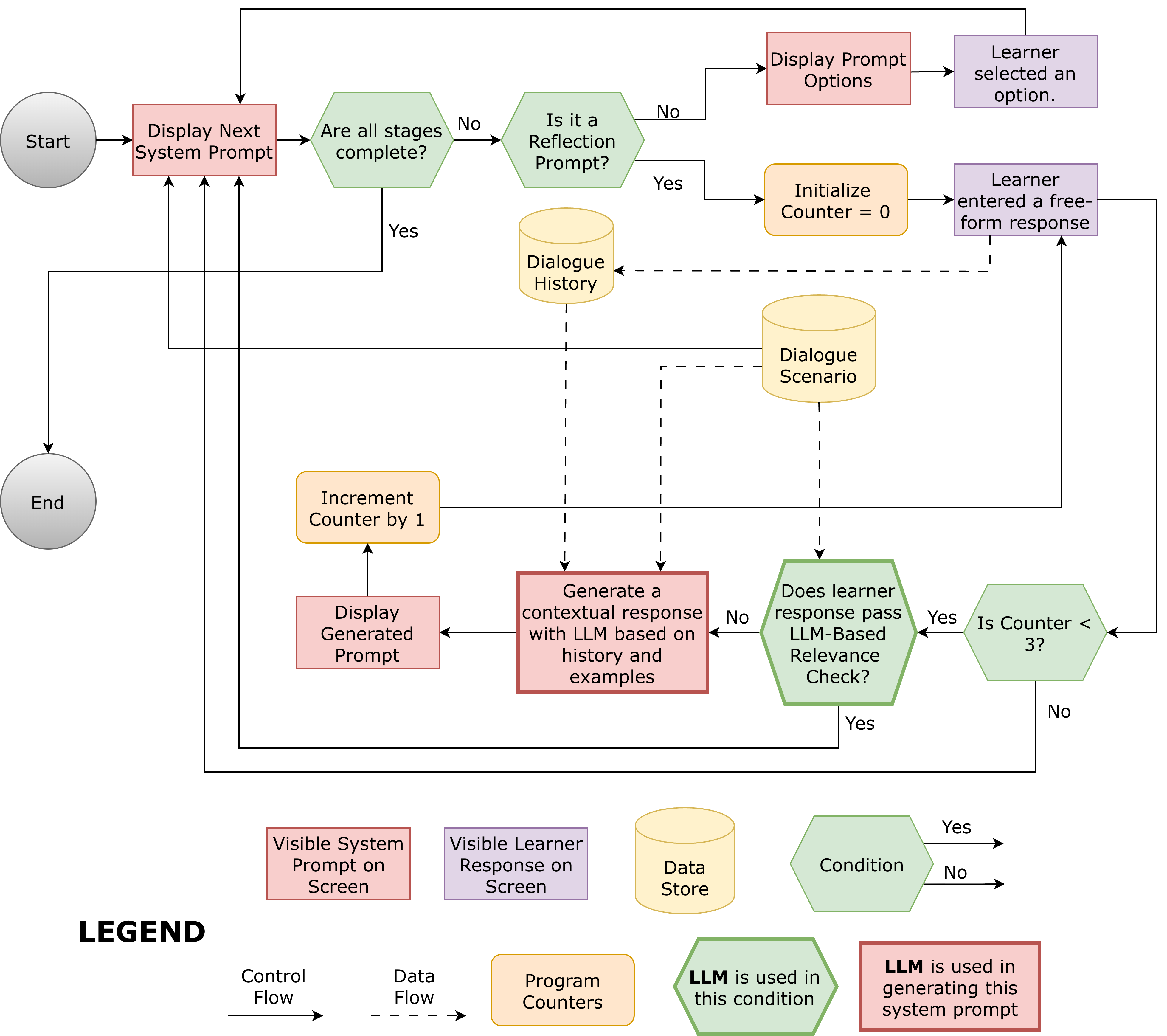}
    \caption{This diagram depicts our hybrid dialogue system architecture modeled as a Finite State Machine and illustrates how logic flows across the rule-based prompts, learner responses, and conditional transitions at each system-learner turn.}
    \Description[The architecture of the hybrid dialogue system is modeled as a finite state machine.]{The figure describes the architecture of our hybrid dialogue system, which is modeled as a Finite State Machine. The logic flow starts with the system displaying a rule-based prompt followed by a learner response. If it is a reflection prompt, then the LLM gets triggered and checks for the response relevance. If the response is relevant, the system moves to the next rule-based response. If the response needs more elaboration, the system triggers LLM for contextual response generation, which is then presented to the user. This cycle continues until the user response is relevant to the input prompt or a counter of three is reached, after which the system defaults to the rule-based prompt.}
    \label{fig:dialogue_arch}
\end{figure*}

%% file: llm_workflow.tex
\begin{figure*}
    \centering
    \includegraphics[width=\linewidth,height=16cm,keepaspectratio]{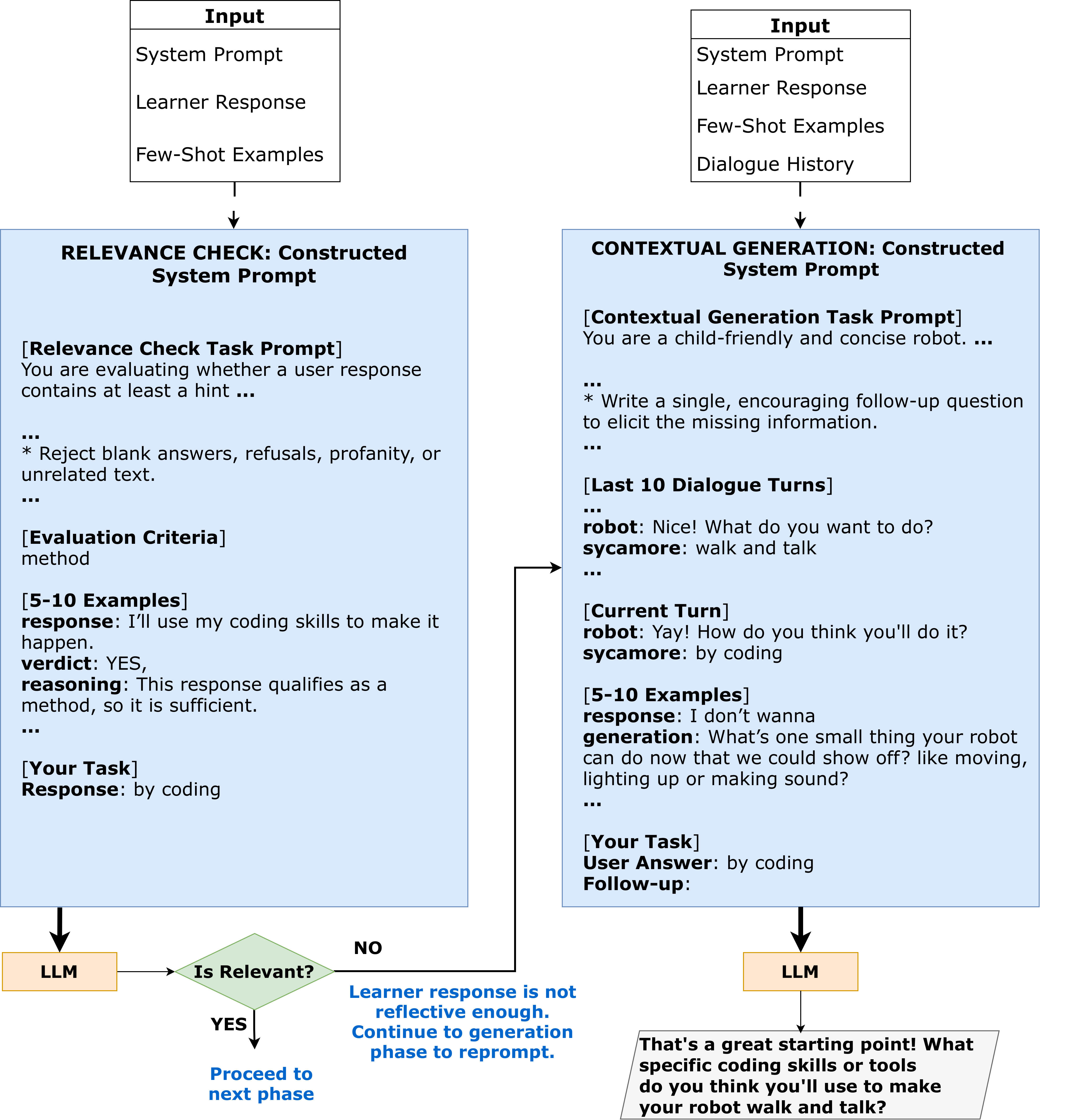}
    \caption{The two stages of LLM in our dialogue system: (1) Relevance Check, where a dynamic prompt with prompt-specific information and few-shot examples determines binary relevance; and (2) Contextual Generation, where, upon a NO in relevance check, a new prompt incorporating prompt-specific information, dialogue history, and few-shot examples guides the LLM in producing a follow-up.}
    \Description[The two stage architecture for LLM check and generation in our dialogue system.]{The figure shows the two stages of LLM in our dialogue system (1) Relevance Check, where a dynamic prompt with prompt-specific information and few-shot examples determines binary relevance; and (2) Contextual Generation, where, upon a NO in relevance check, a new prompt incorporating prompt-specific information, dialogue history, and few-shot examples guides the LLM in producing a follow-up. The figure further shows an example where the learner initially responded ``by coding'' to the system's prompt, which failed the relevance check and triggered the LLM to reprompt with ``That's a great starting point! What specific coding skills or tools do you think you'll use to make your robot walk and talk?''}
    \label{fig:llm-workflow}
\end{figure*}

%% file: chat_interface.tex
\begin{figure}
    \centering
    \includegraphics[width=0.7\linewidth,height=12cm,keepaspectratio]{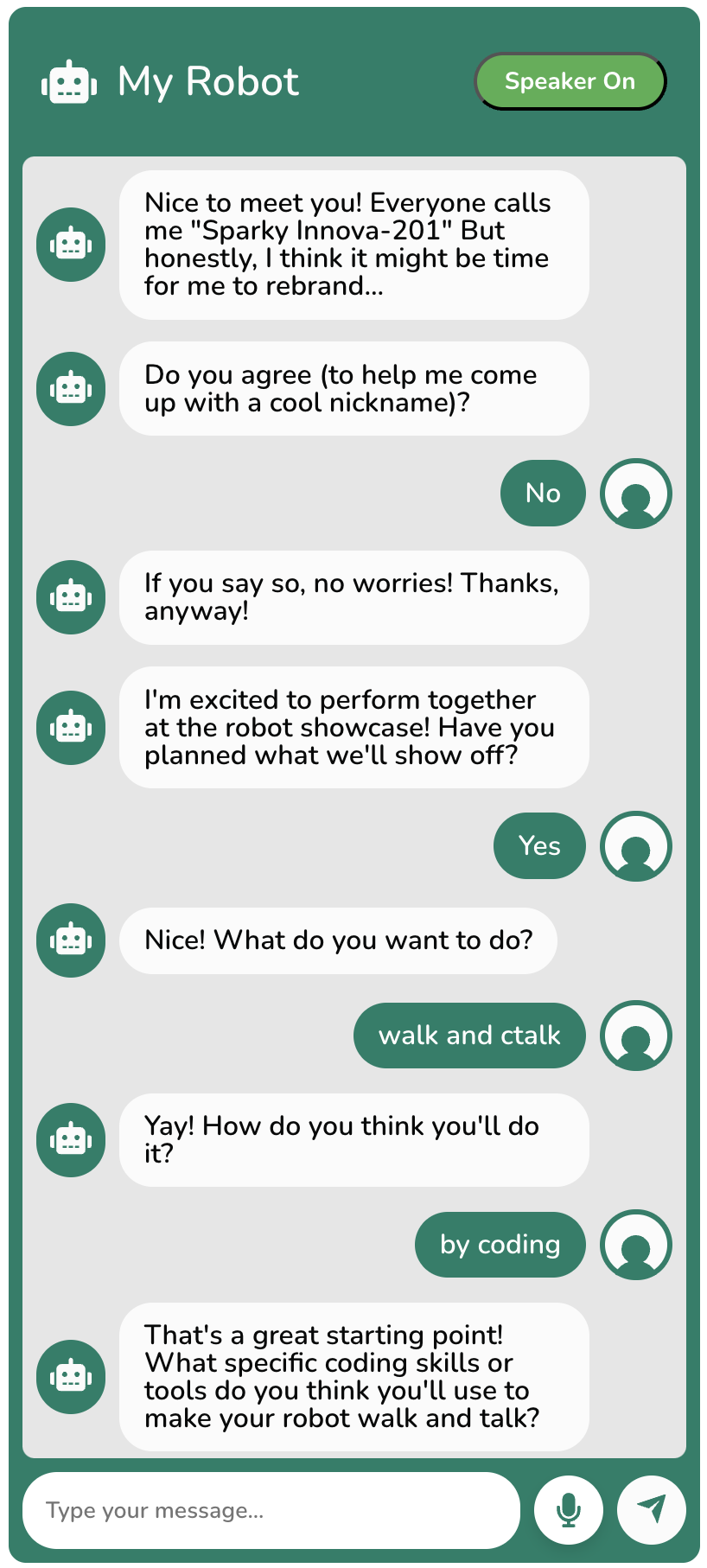}
    \caption{The figure shows the chat widget embedded in our system's frontend programming interface. Learners interact with the system through this widget using text or speech-to-text input, while system responses are delivered as both text and audio. The example displayed shows a dialogue snippet from our camp where the system prompts a learner to reflect on their goals and plans for the robot showcase. The learner responses ``Yes'' and ``No'' are part of a menu-based selection, and ``walk and ctalk'' and ``by coding'' are free-text responses.}
    \Description[The chat widget in our frontend interaction interface.]{The figure shows the chat widget embedded in our system's frontend programming interface. Learners interact with the system through this widget using text or speech-to-text input, while system responses are delivered as both text and audio. There is an example displayed in the figure that shows a dialogue snippet from our camp where the system prompts a learner to reflect on their goals and plans for the robot showcase.}
    \label{fig:chat-interface}
\end{figure}

%% file: codes_table_2.tex
\begin{table*}[t]
\caption{The table shows categories devised (with description and examples) after qualitative coding of the learner dialogue interactions with the system. The categories that emerged were: LLM Triggers, LLM Reprompts, Learner Reflection, Emotion, and Engagement.}
\label{tab:codes}
\begin{tabular}{|l|p{6.9cm}|p{6.9cm}|}
\hline
\textbf{Category}                    & \textbf{Description}                                                                                                                                                                                                                                                                                        & \textbf{Examples}                                                                                                                                                                                                                                                                                                                                                                                                                                              \\ \hline
\textit{\textbf{LLM Triggers}}       & \begin{tabular}[c]{@{}p{6.8cm}@{}}\addlinespace[3pt]Under this category, we specifically looked for instances where LLM was triggered for learner goals, plans, activities, feelings. This also includes changes in these components. We also noted instances where LLM failed to trigger.\end{tabular}                     & \begin{tabular}[c]{@{}p{6.8cm}@{}}\addlinespace[3pt]Robot: No worries! What would you like to do instead? (Goal)\\ Robot: That's a great starting point! What specific coding skills or tools do you think you'll use to make your robot walk and talk? (Plan)\end{tabular}                                                                                                                                                                                                          \\ \hline
\textit{\textbf{LLM Reprompts}}      & \begin{tabular}[c]{@{}p{6.8cm}@{}}\addlinespace[3pt]Under this category, we coded for if the LLM reprompts were either in-context with the conversation and learner inputs or were they out of context. We also looked for instances where the response contained hallucination or bias.\end{tabular}                       & \begin{tabular}[c]{@{}p{6.8cm}@{}}\addlinespace[3pt]Robot: That's alright, not every day is about finishing. Sometimes just trying counts too! Can you think of one thing you started, explored, or even just thought about today? (in-context)\\ Robot: Her hair is an interesting choice. Can you think of a way that working on her hair might be connected to what we talked about earlier, like building a robot that can walk and talk? (out-of-context)\end{tabular} \\ \hline

\textit{\textbf{Learner Reflection}} & \begin{tabular}[c]{@{}p{6.8cm}@{}}\addlinespace[3pt]Under this category, we coded for if the learner reflected on their design goals, plans, actions, or changes to these components or their feelings. We also code for if any of these reflections were elicited due to an LLM reprompt in the previous turn.\end{tabular} & \begin{tabular}[c]{@{}p{6.8cm}@{}}\addlinespace[3pt]Learner: I think I'll do it by putting fur on you (Action)\\ Learner: I coded my robot to move and say word problems (Goal) (LLM)\end{tabular}\\ \hline                                                                                                                      
\textit{\textbf{Emotion}}            & \begin{tabular}[c]{@{}p{6.8cm}@{}}\addlinespace[3pt]Under this category, we coded for explicit mentions of positive sentiments (or logical reasoning steps) and negative sentiments. The default code under this category was neutral.\end{tabular}                                                                            & \begin{tabular}[c]{@{}p{6.8cm}@{}}\addlinespace[3pt]Learner: because i love robots (positive)\\ Learner: leaveeeeeee meeeeee alone (negative)\\ Learner: make you move (neutral)\end{tabular}\\ \hline                                                                                                                
\textit{\textbf{Engagement}}         & \begin{tabular}[c]{@{}p{6.8cm}@{}}\addlinespace[3pt]We this category, we coded for explicit moments which defined high-engagements (providing elaborate response) or low-engagements (mentioning the intention not to talk, or a continuous sequence of  ``no'' in response)\end{tabular}                                   & \begin{tabular}[c]{@{}p{6.8cm}@{}}\addlinespace[3pt]Learner: ok but i have to go bye (low-engagement)\\ Learner: move right for 3 secs and then you'll spin and dance just move your arms (high-engagement)\end{tabular}                                                                                                                                                                                                                                                              \\ \hline
\end{tabular}
\end{table*}

%% file: summary_quant_descriptives_table.tex
\begin{table}
  \centering
  \caption{Descriptive statistics of dialogue sessions ($N=9$ participants). The table details session duration, the breakdown of learner inputs (open-ended vs. option-based), turns, and word counts, and LLM trigger frequency.}
  \label{tab:descriptive_stats}
  \begin{tabular}{lp{3.5cm}cp{1.5cm}}
    \toprule
    \textbf{Category} & \textbf{Metric} & \textbf{Total} & \textbf{Mean (SD)} \\
    \midrule
    \multicolumn{4}{l}{\textit{\textbf{1. Session Overview}}} \\
    & Dialogue Turns & 357 & 39.66 (6.46) \\
    & Session Duration (min) & -- & 6.00 (2.00) \\
    \midrule
    \multicolumn{4}{l}{\textit{\textbf{2. Turn-Level Metrics}}} \\
    & System Turns & 213 & 23.67 (3.64) \\
    & Learner Turns & 144 & 16.00 (2.91) \\
    & \hspace{3mm}\textit{-- Open-Ended} & 88 & 9.77 (2.77) \\
    & \hspace{3mm}\textit{-- Option-Based} & 56 & 6.22 (0.44)\\
    \midrule
    \multicolumn{4}{l}{\textit{\textbf{3. Word Counts}}} \\
    & System Words per Turn & -- & 14.41 (1.06) \\
    & Learner Words per Turn* & -- & 3.89 (1.75) \\
    \midrule
    \multicolumn{4}{l}{\textit{\textbf{4. LLM Triggers}}} \\
    & LLM Triggers per Session & -- & 2.66 (2.74) \\
    & Follow-up Triggers per Session & -- & 1.14 (0.89) \\
    \bottomrule
    \multicolumn{4}{p{8cm}}{\footnotesize{\textit{*Learner word counts calculated on Open-Ended responses only. }}}
  \end{tabular}
\end{table}

%% file: findings_summary_codes_table.tex
\begin{table*}[ht]
  \centering
  \caption{Analysis of system performance for response verbosity, relevance check quality, prompt content, and quality.}
  \label{tab:llm_evaluation}
  \begin{tabular}{lp{9.5cm}l}
    \toprule
    \textbf{Evaluation} & \textbf{Metric} & \textbf{Count / Mean (SD)} \\
    \midrule
    \multicolumn{3}{l}{\textit{\textbf{1. Response Verbosity}}} \\
    & Learner response word count before LLM prompt or re-prompt & 2.59 (1.13) \\
    & Immediate learner response word count post LLM prompt or re-prompt & 3.55 (3.07) \\
    & Expansion Factor b/w post prompt (or re-prompt) and pre prompt lengths & 1.75$\times$ (3.27) \\
    \midrule
    \multicolumn{3}{l}{\textit{\textbf{2. Relevance Check Quality}}} \\
    & LLM correctly triggered (TP) & 23 \\
    & LLM triggered erroneously (FP) & 1 \\
    & LLM correctly remained silent (TN) & 55 \\
    & LLM failed to probe insufficient response (FN) & 12 \\
    \midrule
    \multicolumn{3}{l}{\textit{\textbf{3. LLM Trigger Content ($N=24$)}}} \\
    & Triggered for learner goals/plans/activities & 17 \\
    & Triggered for design changes & 5 \\
    & Triggered for feelings/milestones & 2 \\
    \midrule
    \multicolumn{3}{l}{\textit{\textbf{4. Generation Quality ($N=24$)}}} \\
    & Prompts that elicited elaboration  & 9 \\
    & Prompts that did not elicit elaboration & 15 \\
    & \hspace{3mm}\textit{-- Contextually misaligned} & 4 \\
    & \hspace{3mm}\textit{-- Affectively misaligned} & 11 \\
    \bottomrule
  \end{tabular}
\end{table*}

%% file: rule-based-dialogue.tex
\begin{figure*}
    \centering
    \includegraphics[width=\textwidth,keepaspectratio]{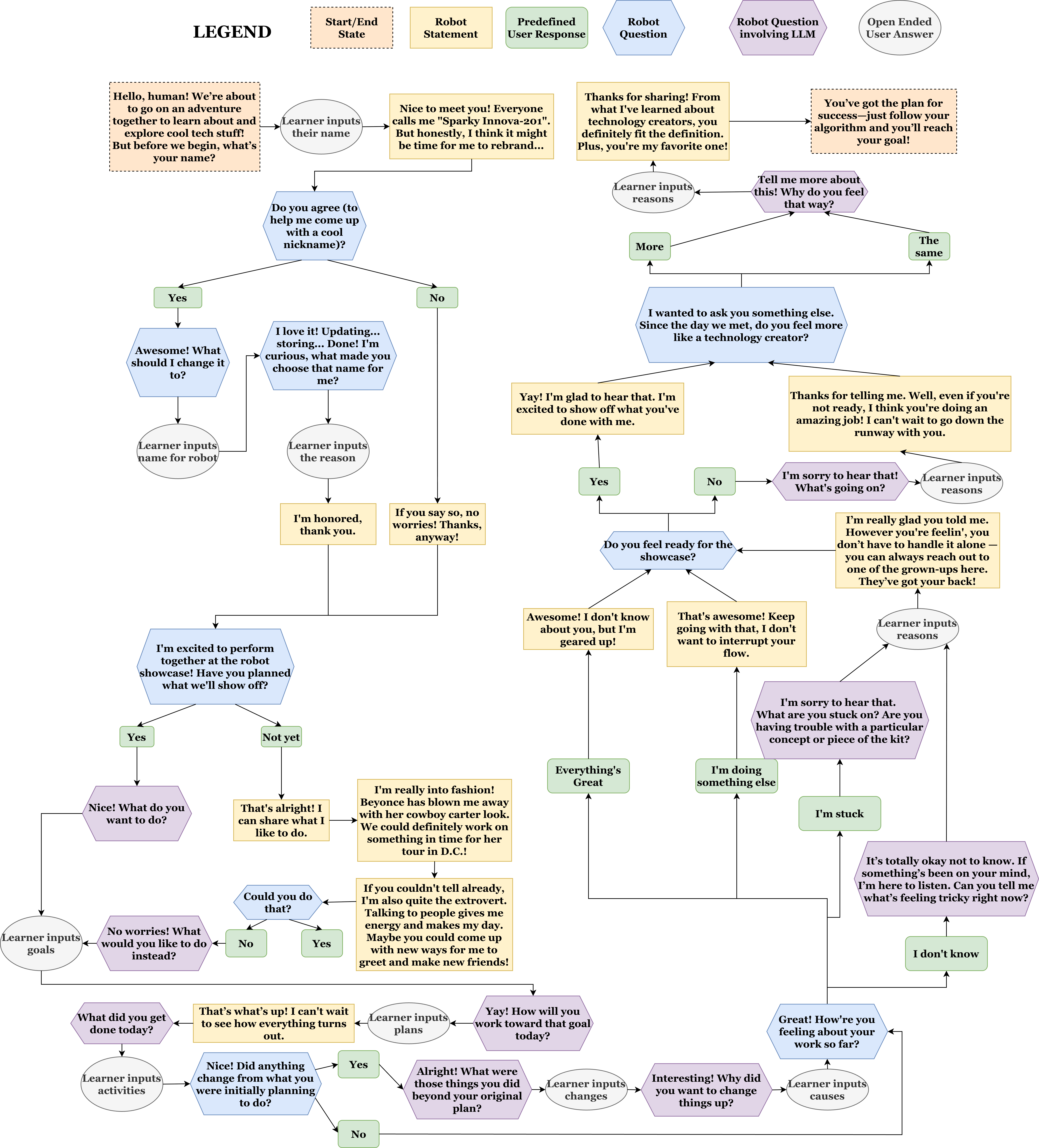}
    \caption{The rule-based prompts for the dialogue system.}
    \Description[A large flow-chart with the rule-based system prompts.]{The figure displays how the dialogues flow across different prompts in the rule-based design. These prompts were defined to know more about learners' goals, plans, actions, and feelings throughout the camp. There is a broad sequence of states containing prompts for building rapport, revisiting goals, plans, and current activities, noting goal or plan changes, explaining causes for changes, feelings about designs for the upcoming showcase, and reflecting on identity as technology creators, each implemented as a node in the state machine.}
    \label{fig:rule-based}
\end{figure*}